\documentclass[12pt,english,sort&compress]{elsarticle}
\usepackage[T1]{fontenc}
\usepackage[latin9]{inputenc}
\usepackage{geometry}
\usepackage{color}
\usepackage{units}
\usepackage{amsbsy}
\usepackage{graphicx}
\usepackage{esint}
\usepackage{babel}
\begin{document}

\title{An energy- and charge-conserving, nonlinearly implicit, electromagnetic
1D-3V Vlasov-Darwin particle-in-cell algorithm}

\author[lanl]{G. Chen\corref{cor1}}

\ead{gchen@lanl.gov}

\author[lanl]{L. Chacón}

\cortext[cor1]{Corresponding author}

\address[lanl]{Los Alamos National Laboratory, Los Alamos, NM 87545}

\address{}
\begin{abstract}
A recent proof-of-principle study proposes a nonlinear electrostatic
implicit particle-in-cell (PIC) algorithm in one dimension (Chen,
Chacón, Barnes, \textit{J. Comput. Phys.} \textbf{230} (2011) 7018).
The algorithm employs a kinetically enslaved Jacobian-free Newton-Krylov
(JFNK) method, and conserves energy and charge to numerical round-off.
In this study, we generalize the method to electromagnetic simulations
in 1D using the Darwin approximation of Maxwell's equations, which
avoids radiative aliasing noise issues by ordering out the light wave.
An implicit, orbit-averaged time-space-centered finite difference
scheme is applied to both the 1D Darwin field equations (in potential
form) and the 1D-3V particle orbit equations to produce a discrete
system that remains exactly charge- and energy-conserving. Furthermore,
enabled by the implicit Darwin equations, exact conservation of the
canonical momentum per particle in any ignorable direction is enforced
via a suitable scattering rule for the magnetic field. Several 1D
numerical experiments demonstrate the accuracy and the conservation
properties of the algorithm.
\end{abstract}
\maketitle

\section{Introdution}

The electromagnetic (EM) Particle-in-cell (PIC) method solves Vlasov-Maxwell's
equations for kinetic plasma simulations \citep{birdsall-langdon,hockneyeastwood}.
In the standard approach, Maxwell's equations are solved on a grid,
and the Vlasov equation is solved by method of characteristics using
a large number of particles, from which the evolution of the probability
distribution function (PDF) is obtained. The field-PDF description
is tightly coupled. Maxwell\textquoteright{}s equations (or a subset
thereof) are driven by moments of the PDF such as charge density and/or
current density. The PDF, on the other hand, follows a hyperbolic
equation in phase space, whose characteristics are determined by the
fields self-consistently.

To date, most PIC methods employ explicit time-stepping (e.g. leapfrog
scheme), which can be very inefficient for long-time, large spatial
scale simulations. The algorithmic inefficiency of standard explicit
PIC is rooted in the presence of numerical stability constraints,
which force both a minimum\textcolor{black}{{} grid-size (due to the
so-called finite-grid instability \citep{birdsall-langdon,hockneyeastwood},
which requires resolu}tion of the smallest Debye length)\textcolor{black}{{}
and a very small timestep (due to the well-known CFL constraint in
the general electromagnetic case, $c\Delta t<\Delta x$, where $c$
is speed of light, and $\Delta t$ and $\Delta x$ are the timestep
and grid-size, respectively)}. Furthermore, numerical heating due
to the lack of exact discrete energy conservation \textcolor{black}{\citep{birdsall-langdon,hockneyeastwood}
compromises the accuracy of explicit PIC simulations over long time
scales}, particularly for realistic ion-to-electron mass ratios\textcolor{black}{.}
In the electromagnetic context, the accuracy issue is aggravated further
by the presence of electromagnetic waves, which can be either unstable
\citep{godfrey1974numerical} or noisily excited to high levels \citep{langdon1972some}.

Implicit methods, however, can free the PIC approach from numerical
stability constraints, and thus have the potential of much improved
algorithmic efficiency. This realization drove the exploration of
implicit PIC starting in the 1980s \citep{mason-jcp-81-im_pic,denavit-jcp-81-im_pic,brackbill-forslund,celeste1d,friedman-cppcf-81-di_pic,cohen-jcp-82-ipic,langdon-jcp-83-di_pic,barnes-jcp-83-di_pic,brackbill-mts-85,langdon1985multiple,cohen1986multiple,mason1987electromagnetic,hewett-jcp-87-di_pic,friedman1990second,kamimura1992implicit,celeste3d,gibbons1995darwin}.
These studies explored the viability of an implicit PIC formulation
and its accuracy properties, and resulted in important developments
such as the implicit-moment method \citep{mason-jcp-81-im_pic,denavit-jcp-81-im_pic,brackbill-forslund,brackbill-mts-85,celeste1d,celeste3d}
and the direct-implicit method \citep{friedman-cppcf-81-di_pic,langdon-jcp-83-di_pic,langdon1985multiple,hewett-jcp-87-di_pic,kamimura1992implicit}.
However, limitations of the solver technology at the time forced early
implicit PIC practitioners to rely on approximations such as linearization
and lagging, which did not respect the strong field-particle coupling.
These numerical approximations produced energy conservation errors
that could result in significant plasma self-heating or self-cooling
\citep{cohen-jcp-89-ipic_perf}.

Fully implicit algorithms hold the promise of overcoming some of the
difficulties of explicit and seim-implicit EM-PIC schemes. Some of
these advantages were demonstrated in Ref. \citep{markidis2011energy},
where an energy-conserving fully implicit Vlasov-Maxwell EM-PIC scheme
was proposed. However, it was shown in the reference that the approach
suffered from radiative aliasing noise, which obscures physical signals
as errors accumulate in time. The radiative noise disappeared by introducing
some numerical damping in the discretization, but this in turn destroyed
the exact energy conservation property.

In non-relativistic applications, radiative aliasing noise can be
eliminated by ordering out light waves from Maxwell's equations to
arrive to the so-called Darwin model \citep{darwin1920li,hasegawa1968one,kaufman1971darwin,krause2007unified}.
The Darwin field equations are no longer hyperbolic, but elliptic,
rendering explicit time integration schemes unconditionally unstable
\citep{nielson-darwin-76}. \textcolor{black}{Nielson and Lewis }\citep{nielson-darwin-76}
introduced semi-implicit schemes to advance the Darwin-PIC system,
which have become the standard for later development and applications
of plasma Darwin-PIC simulations (see Refs. \citep{busnardo1977self,byers1978hybrid,hewett1994low,gibbons1995darwin,sonnendrucker1995finite,lee2001nonlinear,taguchi2004study,borodachev2006numerical,eremin2013simulations}
and references therein). Nevertheless, the resulting field equations,
in either Hamiltonian or Lagrangian form, are complicated and difficult
to solve, especially when non-periodic boundary conditions are employed
\citep{weitzner1989boundary,degond1992analysis,hewett1994low,sonnendrucker1995finite},
and feature no exact conservation properties (e.g., local charge,
total energy, or total momentum).

In contrast to earlier implicit Darwin-PIC studies, our focus here
is on fully implicit, fully nonlinear PIC algorithms. We build upon
recent developments in fully implicit electrostatic \citep{chen-jcp-11-ipic,taitano-sisc-13-ipic}
and electromagnetic \citep{markidis2011energy} PIC algorithms, which
enforce tight nonlinear convergence between particles and fields at
every timestep. Their fully implicit character enables exact discrete
conservation properties, such as energy and charge conservation, which
are attractive for long-time simulations. 

The purpose of this study is to demonstrate a fully implicit scheme
for the Dawin model that conserves energy and charge exactly in a
discrete setting, without suffering from enhanced radiative aliasing
noise \citep{markidis2011energy}. The Darwin equations are solved
in potential form in a one-dimensional (1D) periodic system \citep{hasegawa1968one}
using a Jacobian-free Newton-Krylov (JFNK) solver \citep{chen-jcp-11-ipic,markidis2011energy}.
Particle orbit equations involving three velocity components and one
position are solved implicitly with particle sub-stepping and orbit-averaging
\citep{cohen1985orbit,chen-jcp-11-ipic}. Special care is taken when
scattering the magnetic field to the particles, so that the particle
canonical momentum in any ignorable direction is conserved exactly. 

The aim and intent of this study resonates strongly with an earlier
implementation of the 1D-3V Darwin-PIC model by Hasegawa et al. \citep{hasegawa1968one}.
In this reference, the authors prove conservation theorems for local
charge, global energy, and particle canonical momenta in a continuum-time
Klimontovich representation of the plasma system. The study in the
present paper goes beyond Hasegawa and co-author's in that the conservation
theorems are proved in a discrete setting. The fully implicit character
of our implementation turns out to be key to realize these discrete
conservation properties.

The rest of the paper is organized as follows. Section \ref{sec:Electromagnetic-Vlasov-Darwin}
introduces our formulation for the general Vlasov-Darwin model and
its favorable properties. The model is reduced to 1D-3V and discretized
with an implicit particle-based central-difference scheme in Sec.
\ref{sec:1d-Darwin-PIC}, where we review our charge-conserving particle-moving
strategy, and prove theorems for the exact conservation of global
energy and particle canonical momenta in a discrete setting. Numerical
examples demonstrating the properties of the algorithm are presented
in Sec. \ref{sec:numerical-tests}. Finally, we conclude in Sec. \ref{sec:conclusions}.

\section{Electromagnetic Vlasov-Darwin model\label{sec:Electromagnetic-Vlasov-Darwin}}

The general Vlasov-Darwin model for a collisionless electromagnetic
plasma reads \citep{nielson-darwin-76,Hewett-ssr-85-darwin,degond1992analysis,raviart1996hierarchy,krause2007unified}:
\begin{eqnarray}
\partial_{t}f_{\alpha}+\mathbf{v}\cdot\nabla f_{\alpha}+\frac{q_{\alpha}}{m_{\alpha}}(\mathbf{E}+\mathbf{v}\times\mathbf{B})\cdot\nabla_{v}f_{\alpha} & = & 0,\label{eq:Vlasov}\\
\frac{1}{\mu_{0}}\nabla\times\nabla\times\mathbf{A}=-\frac{1}{\mu_{0}}\nabla^{2}\mathbf{A} & = & \mathbf{j}-\epsilon_{0}\partial_{t}\nabla\phi,\label{eq:Darwin-Ampere}\\
\epsilon_{0}\nabla^{2}\phi & = & -\rho,\label{eq:Ampere-phi}\\
\nabla\cdot\mathbf{A} & = & 0,\label{eq:div-A}
\end{eqnarray}
where $f_{\alpha}(\mathbf{r},\mathbf{v})$ is the particle distribution
function of species $\alpha$ in phase space, $q_{\alpha}$ and $m_{\alpha}$
are the species charge and mass respectively, $\epsilon_{0}$ and
$\mu_{0}$ are the vacuum permittivity and permeability respectively,
$\phi$ and $\mathbf{A}$ are the self-consistent electric and vector
potential respectively. Unlike Maxwell's equations, the Darwin model
does not feature Gauge invariance, and only the Coulomb gauge is physically
acceptable (to enforce charge conservation \citep{krause2007unified,hasegawa1968one},
as discussed below). The electric and magnetic fields are defined
uniquely from $\phi,$ $\mathbf{A}$ as:
\begin{equation}
\mathbf{E}=-\nabla\phi-\partial_{t}\mathbf{A}\,\,;\,\,\mathbf{B}=\nabla\times\mathbf{A}.\label{eq:E-def-potentials}
\end{equation}
The Darwin equations are driven by the plasma current density $\mathbf{j}=\sum_{\alpha}q_{\alpha}\int f_{\alpha}\mathbf{v}d\mathbf{v}$
and charge density $\rho=\sum_{\alpha}q_{\alpha}\int f_{\alpha}d\mathbf{v}$.

The Vlasov-Darwin model in Eqs. \ref{eq:Vlasov}-\ref{eq:div-A} features
two involutions, Poisson's equation and the solenoidal constraint
of the vector potential. However, this system is overdetermined, and
can be formulated much more succintly, as we shall see. We begin by
realizing that the local charge conservation equation,
\begin{equation}
\partial_{t}\rho+\nabla\cdot\mathbf{j}=0,\label{eq:charge-cons}
\end{equation}
can be derived independently from both the Vlasov equation (Eq. \ref{eq:Vlasov})
and the Darwin electromagnetic model (Eqs. \ref{eq:Darwin-Ampere}-\ref{eq:div-A}).
From the Vlasov equation, Eq. \ref{eq:charge-cons} follows by taking
its zeroth velocity moment for all species, and then adding them up
according to the definition of the charge density. From the Darwin
equations, the charge conservation equation follows by taking the
divergence of Eq. \ref{eq:Darwin-Ampere}, and using Eq. \ref{eq:Ampere-phi}.

This redundancy in the model can be exploited to formulate a minimal
set of Darwin equations such that, together with the Vlasov equation,
the two involutions are implied in the formulation, and do not need
to be enforced explicitly. This minimal Darwin model is comprised
of two equations. The first equation is the vector Laplacian form
of Eq. \ref{eq:Darwin-Ampere}. The second equation is found by taking
the divergence of Eq. \ref{eq:Darwin-Ampere} and using Eq. \ref{eq:div-A}.
Thus, our final set of Vlasov-Darwin equations reads: 
\begin{eqnarray}
\partial_{t}f_{\alpha}+\mathbf{v}\cdot\nabla f_{\alpha}+\frac{q_{\alpha}}{m_{\alpha}}(\mathbf{E}+\mathbf{v}\times\mathbf{B})\cdot\nabla_{v}f_{\alpha} & = & 0,\label{eq:Darwin-Vlasov}\\
-\frac{1}{\mu_{0}}\nabla^{2}\mathbf{A} & = & \mathbf{j}-\epsilon_{0}\partial_{t}\nabla\phi,\label{eq:Darwin-ampere-2}\\
\epsilon_{0}\partial_{t}\nabla^{2}\phi & = & \nabla\cdot\mathbf{j}.\label{eq:Darwin-Poisson}
\end{eqnarray}
These three equations, together with the local charge conservation
equation (which is implicit in Eq. \ref{eq:Darwin-Vlasov}), imply
the involutions (Eqs. \ref{eq:Ampere-phi}, \ref{eq:div-A}). In particular,
Poisson's equation (Eq. \ref{eq:Ampere-phi}) is implied by Eq. \ref{eq:Darwin-Poisson}
and the charge conservation equation. The solenoidal constraint is
implied as well. This is seen by taking the divergence of Eq. \ref{eq:Darwin-ampere-2}
and using Eq. \ref{eq:Darwin-Poisson}, to find:
\[
\nabla^{2}\nabla\cdot\mathbf{A}=0,
\]
from which, with appropriate boundary conditions, Eq. \ref{eq:div-A}
follows. The boundary conditions must be consistent with $\nabla\cdot\mathbf{A}=0$
at the boundary \citep{hasegawa1968one} (i.e., must enforce continuity
of the normal component of the vector potential at the boundary).

Equations \ref{eq:Darwin-Vlasov}, \ref{eq:Darwin-ampere-2}, and
\ref{eq:Darwin-Poisson} constitute the minimal Vlasov-Darwin set
of choice in this study. We emphasize that the main advantage of this
set is that the two involutions (Poisson's equation and the solenoidal
constraint of $\mathbf{A}$) are implied, and thus do not need to
be enforced or solved explicitly. This property, when implemented
discretely, will be most advantageous, as ensuring (or avoiding) Eq.
\ref{eq:div-A} has turned out to be one the most difficult implementation
roadblocks of the Darwin approximation in multiple dimensions \citep{nielson-darwin-76,hewett1994low}.
This, however, will require a very careful discrete treatment, and
in particular one that strictly conserves local charge.

\section{One-dimensional implicit particle-based discretization of the Vlasov-Darwin
model\label{sec:1d-Darwin-PIC}}

In the remainder of this study, we specialize the Vlasov-Darwin equations
to one spatial dimension and three velocity dimensions (1D-3V) in
Cartesian geometry, as follows.%
\footnote{We should point out that enforcing the solenoidal involution is trivial
in this reduced dimensionality context, and therefore the point outlined
in the previous section about implied involutions is not so critical
for $\nabla\cdot\mathbf{A}=0$. However, it will be key in multiple
dimensions. We will comment on the extension of the current approach
to multiple dimensions later in this paper.%
} We consider a 1D periodic system, with $\partial_{y}=\partial_{z}=0$,
for which the Darwin model reduces to: 
\begin{eqnarray}
\epsilon_{0}\partial_{t}E_{x}+j_{x} & = & \left\langle j_{x}\right\rangle ,\label{eq:Ampere-1d}\\
\frac{1}{\mu_{0}}\partial_{x}^{2}A_{y,z}+j_{y,z} & = & \left\langle j_{y,z}\right\rangle ,\label{eq:Ayz-1d}
\end{eqnarray}
where $E_{x}=-\partial\phi/\partial x$, and the terms on the right
hand side are the spatial average of the current densities, e.g. $\left\langle j_{x}\right\rangle =\int j_{x}dx/\int dx$.
These are necessary in a periodic system to enforce periodicity of
the fields \citep{hasegawa1968one,chen-jcp-11-ipic}. The inductive
electric field is determined from the vector potential as:
\begin{equation}
E_{y,z}=-\partial_{t}A_{y,z}.\label{eq:Eyz-induced}
\end{equation}
The magnetic field is determined from the vector potential as:
\begin{equation}
B_{x}=B_{0x}\,\,;\,\, B_{y}=B_{0y}-\partial_{x}A_{z}\,\,;\,\, B_{z}=B_{0z}+\partial_{x}A_{y},\label{eq:B-def}
\end{equation}
where $(B_{0x},B_{0y},B_{0z})$ is a prescribed equilibrium magnetic
field. 

Discretizing the 1D equations with central difference in time gives
for the electric field components:
\begin{eqnarray}
\epsilon_{0}\frac{E_{x,i+\nicefrac{1}{2}}^{n+1}-E_{x,i+\nicefrac{1}{2}}^{n}}{\Delta t}+j_{x,i+\nicefrac{1}{2}}^{n+\nicefrac{1}{2}} & = & \left\langle j_{x}\right\rangle ,\label{eq:Ex}\\
E_{y,i}^{n+\nicefrac{1}{2}} & = & -\frac{A_{y,i}^{n+1}-A_{y,i}^{n}}{\Delta t},\label{eq:Ey}\\
E_{z,i}^{n+\nicefrac{1}{2}} & = & -\frac{A_{z,i}^{n+1}-A_{z,i}^{n}}{\Delta t},\label{eq:Ez}
\end{eqnarray}
where the superscript $n$ denotes the time level at $n\Delta t$,
the subscript $i$ denotes the mesh point at $i\Delta x$, and $\Delta t$
and $\Delta x$ are time and spatial mesh intervals respectively for
the field equations. The vector potential components are found from:
\begin{eqnarray}
\frac{1}{\mu_{0}}\left.\partial_{x}^{2}\frac{A_{y}^{n+1}+A_{y}^{n}}{2}\right|_{i}+j_{y,i}^{n+\nicefrac{1}{2}} & = & \left\langle j_{y}\right\rangle ,\label{eq:Ay}\\
\frac{1}{\mu_{0}}\left.\partial_{x}^{2}\frac{A_{z}^{n+1}+A_{z}^{n}}{2}\right|_{i}+j_{z,i}^{n+\nicefrac{1}{2}} & = & \left\langle j_{z}\right\rangle .\label{eq:Az}
\end{eqnarray}
Note that the vector components along ignorable directions are defined
at the integer spatial mesh points, while the $x$-components are
defined at the half spatial mesh points. The $\partial_{x}^{2}$ is
discretized using a standard central-difference formula, e.g. $\left.\partial_{x}^{2}A\right|_{i}=(A_{i+1}-2A_{i}+A_{i-1})/\Delta x$.
Similarly, we obtain the magnetic field components as: 
\begin{equation}
B_{y,i+\nicefrac{1}{2}}=B_{0y}-\frac{A_{z,i+1}-A_{z,i}}{\Delta x}\,\,;\,\, B_{z,i+\nicefrac{1}{2}}=B_{0z}+\frac{A_{y,i+1}-A_{y,i}}{\Delta x}.\label{eq:Byz-def}
\end{equation}
The current components are gathered from particles to ensure charge
and energy conservation, as described later in this section.

Particle quantities are evolved from the 1D-3V particle equations
of motion:
\begin{eqnarray}
\partial_{t}x_{p} & = & v_{x,p},\label{eq:x_p}\\
\partial_{t}v_{x,p} & = & \frac{q_{p}}{m_{p}}(E_{x,p}+v_{p,y}B_{z,p}-v_{p,z}B_{y,p}),\label{eq:vx_p}\\
\partial_{t}v_{y,p} & = & \frac{q_{p}}{m_{p}}(E_{y,p}+v_{p,z}B_{x,p}-v_{p,x}B_{z,p}),\label{eq:vy_p}\\
\partial_{t}v_{z,p} & = & \frac{q_{p}}{m_{p}}(E_{z,p}+v_{p,x}B_{y,p}-v_{p,y}B_{x,p}).\label{eq:vz_p}
\end{eqnarray}
As in Ref. \citep{chen-jcp-11-ipic}, these equations are discretized
using a sub-stepped Crank-Nicolson scheme:
\begin{eqnarray}
\frac{x_{p}^{\nu+1}-x_{p}^{\nu}}{\Delta\tau^{\nu}} & = & v_{x,p}^{\nu+\nicefrac{1}{2}},\label{eq:x_p-1}\\
\frac{v_{x,p}^{\nu+1}-v_{x,p}^{\nu}}{\Delta\tau^{\nu}} & = & \frac{q_{p}}{m_{p}}(E_{x,p}^{\nu+\nicefrac{1}{2}}+v_{p,y}^{\nu+\nicefrac{1}{2}}B_{z,p}^{\nu+\nicefrac{1}{2}}-v_{p,z}^{\nu+\nicefrac{1}{2}}B_{y,p}^{\nu+\nicefrac{1}{2}}),\label{eq:vx_p-1}\\
\frac{v_{y,p}^{\nu+1}-v_{y,p}^{\nu}}{\Delta\tau^{\nu}} & = & \frac{q_{p}}{m_{p}}(E_{y,p}^{\nu+\nicefrac{1}{2}}+v_{p,z}^{\nu+\nicefrac{1}{2}}B_{x,p}^{\nu+\nicefrac{1}{2}}-v_{p,x}^{\nu+\nicefrac{1}{2}}B_{z,p}^{\nu+\nicefrac{1}{2}}),\label{eq:vy_p-1}\\
\frac{v_{z,p}^{\nu+1}-v_{z,p}^{\nu}}{\Delta\tau^{\nu}} & = & \frac{q_{p}}{m_{p}}(E_{z,p}^{\nu+\nicefrac{1}{2}}+v_{p,x}^{\nu+\nicefrac{1}{2}}B_{y,p}^{\nu+\nicefrac{1}{2}}-v_{p,y}^{\nu+\nicefrac{1}{2}}B_{x,p}^{\nu+\nicefrac{1}{2}}),\label{eq:vz_p-1}
\end{eqnarray}
where the substep $\Delta\tau^{\nu}$ satisfies $\sum_{\nu=0}^{N_{\nu}}\Delta\tau^{\nu}=\Delta t$,
with $\nu$ denoting the substep and $N_{\nu}$ the number of substeps.
Following earlier studies \citep{chen2013analytical}, the time step
is determined here by a local error estimator $\Delta\tau=0.1min(\omega_{t}^{-1},\omega_{c}^{-1})$,
where $\omega_{t}=\frac{q}{m}|\partial_{x}E|$ is the electrostatic
harmonic frequency, and $\omega_{c}=\frac{q}{m}B$ is the gyrofrequency.

The scatter of the electric field to particles is defined as:
\begin{eqnarray}
E_{x,p}^{\nu+\nicefrac{1}{2}} & = & \sum_{i}\frac{E_{x,i+\nicefrac{1}{2}}^{n+1}+E_{x,i+\nicefrac{1}{2}}^{n}}{2}S_{m}(x_{p}^{\nu+\nicefrac{1}{2}}-x_{i+\nicefrac{1}{2}}),\label{eq:Ex_p}\\
E_{y,p}^{\nu+\nicefrac{1}{2}} & = & -\sum_{i}\frac{A_{y,i}^{n+1}-A_{y,i}^{n}}{\Delta t}S_{l}(x_{p}^{\nu+\nicefrac{1}{2}}-x_{i}),\label{eq:Ey_p}\\
E_{z,p}^{\nu+\nicefrac{1}{2}} & = & -\sum_{i}\frac{A_{z,i}^{n+1}-A_{z,i}^{n}}{\Delta t}S_{l}(x_{p}^{\nu+\nicefrac{1}{2}}-x_{i}),\label{eq:Ez_p}
\end{eqnarray}
where we have assumed that the electric field varies slowly during
the timestep $\Delta t$ \citep{chen-jcp-11-ipic}. Here, $S_{m}$
is the B-spline of order $m$. We will be using $m=1$ and $l=2$
throughout this study. The latter ensures a linear interpolation of
the magnetic field to the particles. The scattering of the magnetic
field components to the particles will be determined such that the
particle canonical momentum in both $y$ and $z$ directions is conserved
exactly, and will be discussed later in this paper. 

The current components needed in Eqs. \ref{eq:Ex}, \ref{eq:Ay},
and \ref{eq:Az} are found from particle quantities as:
\begin{eqnarray}
\bar{j}_{x,i+\nicefrac{1}{2}}^{n+\nicefrac{1}{2}} & = & \frac{1}{\Delta t\Delta x}\sum_{p}\sum_{\nu}q_{p}v_{p,x}^{\nu+\nicefrac{1}{2}}S_{m}(x_{p}^{\nu+\nicefrac{1}{2}}-x_{i+\nicefrac{1}{2}})\Delta\tau^{\nu},\label{eq:jx-avg}\\
\bar{j}_{y,i}^{n+\nicefrac{1}{2}} & = & \frac{1}{\Delta t\Delta x}\sum_{p}\sum_{\nu}q_{p}v_{p,y}^{\nu+\nicefrac{1}{2}}S_{l}(x_{p}^{\nu+\nicefrac{1}{2}}-x_{i})\Delta\tau^{\nu},\label{eq:jy-avg}\\
\bar{j}_{z,i}^{n+\nicefrac{1}{2}} & = & \frac{1}{\Delta t\Delta x}\sum_{p}\sum_{\nu}q_{p}v_{p,z}^{\nu+\nicefrac{1}{2}}S_{l}(x_{p}^{\nu+\nicefrac{1}{2}}-x_{i})\Delta\tau^{\nu},\label{eq:jz-avg}
\end{eqnarray}
where we have added an overbar to denote that the current components
are orbit averaged. Note that $\bar{j}_{y,i}^{n+\nicefrac{1}{2}}$
and $\bar{j}_{z,i}^{n+\nicefrac{1}{2}}$ use a spline order different
from $\bar{j}_{x,i+\nicefrac{1}{2}}^{n+\nicefrac{1}{2}}$ for consistency
with those used by the electric field components (which will in turn
be required for exact energy conservation). Next, we comment on our
procedure to ensure exact charge conservation, and derive energy and
canonical momenta conservation theorems.

\subsection{Charge conservation}

Exact local charge conservation can be ensured kinematically by pushing
particles following the prescription outlined in Ref. \citep{chen-jcp-11-ipic}.
In particular, for $m\leq1$, the continuity equation is satisfied
to numerical round-off whenever particles are forced to land at cell
boundaries along their orbit.

We should note that the use of different spline orders in the current
components in Eqs. \ref{eq:jx-avg}-\ref{eq:jz-avg} does not break
charge conservation, because the current components in the ignorable
directions do not enter the 1D continuity equation. We should also
note that this prescription can be generalized to multiple dimensions
\citep{chen2013analytical}.

\subsection{Energy conservation theorem }

As in earlier studies \citep{hasegawa1968one,chen-jcp-11-ipic,chacon-jcp-12-ipic_curv,chen2013analytical},
we begin by dotting the particle velocity equations, Eqs. \ref{eq:vx_p}-\ref{eq:vz_p},
with the averaged velocity $\mathbf{v}^{\nu+\nicefrac{1}{2}}$, orbit
averaging all substeps, and summing over all particles, to find:
\begin{eqnarray*}
\frac{K^{n+1}-K^{n}}{\Delta t} & = & \sum_{p}\frac{1}{\Delta t}\sum_{\nu}m_{p}\frac{\mathbf{v}_{p}^{\nu+1}+\mathbf{v}_{p}^{\nu}}{2}\cdot\frac{\mathbf{v}_{p}^{\nu+1}-\mathbf{v}_{p}^{\nu}}{\Delta\tau^{\nu}}\Delta\tau^{\nu}=\sum_{p}\frac{1}{\Delta t}\sum_{\nu}q_{p}\left(\mathbf{v}_{p}\cdot\mathbf{E}_{p}\right)^{\nu+\nicefrac{1}{2}}\Delta\tau^{\nu}\\
 & = & \sum_{i}\Delta x\left(E_{x,i+\nicefrac{1}{2}}^{n+\nicefrac{1}{2}}\bar{j}_{x,i+\nicefrac{1}{2}}^{n+\nicefrac{1}{2}}+E_{y,i}^{n+\nicefrac{1}{2}}\bar{j}_{y,i}^{n+\nicefrac{1}{2}}+E_{z,i}^{n+\nicefrac{1}{2}}\bar{j}_{z,i}^{n+\nicefrac{1}{2}}\right),
\end{eqnarray*}
where $K\equiv\sum_{p}\frac{1}{2}m_{p}v_{p}^{2}$ is the total particle
kinetic energy, and we have used Eqs. \ref{eq:Ex_p}-\ref{eq:Ez_p}
and \ref{eq:jx-avg}-\ref{eq:jz-avg}, assuming that the cell width
$\Delta x$ is uniform across the domain. Plugging in Eqs. \ref{eq:Ex}-\ref{eq:Az},
we find:
\begin{eqnarray*}
\sum_{i}\Delta xE_{x,i+\nicefrac{1}{2}}^{n+\nicefrac{1}{2}}\bar{j}_{x,i+\nicefrac{1}{2}}^{n+\nicefrac{1}{2}} & = & -\epsilon_{0}\sum\Delta x\frac{E_{x,i+\nicefrac{1}{2}}^{n+1}+E_{x,i+\nicefrac{1}{2}}^{n}}{2}\frac{E_{x,i+\nicefrac{1}{2}}^{n+1}-E_{x,i+\nicefrac{1}{2}}^{n}}{\Delta t}\\
 & = & -\frac{\epsilon_{0}}{2\Delta t}\sum_{i}\Delta x\left[\left(E_{x,i+\nicefrac{1}{2}}^{n+1}\right)^{2}-\left(E_{x,i+\nicefrac{1}{2}}^{n}\right)^{2}\right]=-\frac{W_{\phi x}^{n+1}-W_{\phi x}^{n}}{\Delta t},\\
\sum_{i}\Delta xE_{y,i}^{n+\nicefrac{1}{2}}\bar{j}_{y,i}^{n+\nicefrac{1}{2}} & = & \frac{1}{\mu_{0}}\sum_{i}\Delta x\left(\frac{A_{y,i}^{n+1}-A_{y,i}^{n}}{\Delta t}\right)\left(\partial_{x}^{2}\frac{A_{y}^{n+1}+A_{y}^{n}}{2}\right)_{i}\\
 & = & -\frac{1}{2\mu_{0}\Delta t}\sum_{i}\Delta x\left[\left(\partial_{x}A_{y}^{n+1}\right)_{i+\nicefrac{1}{2}}^{2}-\left(\partial_{x}A_{y}^{n}\right)_{i+\nicefrac{1}{2}}^{2}\right]=-\frac{W_{Bz}^{n+1}-W_{Bz}^{n}}{\Delta t},\\
\sum_{i}\Delta xE_{z,i}^{n+\nicefrac{1}{2}}\bar{j}_{z,i}^{n+\nicefrac{1}{2}} & = & -\frac{1}{2\mu_{0}\Delta t}\sum_{i}\Delta x\left[\left(\partial_{x}A_{z}^{n+1}\right)_{i+\nicefrac{1}{2}}^{2}-\left(\partial_{x}A_{z}^{n}\right)_{i+\nicefrac{1}{2}}^{2}\right]=-\frac{W_{By}^{n+1}-W_{By}^{n}}{\Delta t}.
\end{eqnarray*}
In these equations, $W_{\phi x}\equiv\frac{\epsilon_{0}}{2}\sum_{i}\Delta xE_{x,i+\nicefrac{1}{2}}^{2}$
is the electrostatic energy, and $W_{By,z}\equiv\frac{1}{2\mu_{0}}\sum_{i}\Delta x\left(B_{y,z}-B_{0y,z}\right)_{i+\nicefrac{1}{2}}^{2}$
is the magnetic energy. Numerical spatial derivatives have been telescoped,
as allowed by a standard central finite differencing of the spatial
second-order derivative. The terms associated with the average currents
in Eqs. \ref{eq:Ex}, \ref{eq:Ay} and \ref{eq:Az} cancel exactly
because $\sum_{i}E_{x,i+\nicefrac{1}{2}}=0$ and $\sum_{i}E_{y,i}=\sum_{i}E_{z,i}=0$.
The former follows from $E_{x}$ being a gradient in a periodic domain.
The latter follow because the average of the corresponding vector
potential component satisfying Eqs. \ref{eq:Ay}, \ref{eq:Az} in
a periodic domain is conserved in time (see App. \ref{app:cons-spatial-average}).
This property transfers to the discrete when the standard discretization
of $\partial_{x}^{2}A_{y,z}$ is used. As a result, $\sum_{i}A_{y,i}^{n+1}=\sum_{i}A_{y,i}^{n}$
and similarly with $A_{z}$. The energy conservation theorem sought
follows:
\begin{equation}
\left(K_{p}+W_{\phi x}+W_{By}+W_{Bz}\right)^{n+1}=\left(K_{p}+W_{\phi x}+W_{By}+W_{Bz}\right)^{n}.\label{eq:disc-totenergy}
\end{equation}

\subsection{Conservation of particle canonical momenta}

One subtlety of the one dimensional electromagnetic system is that
the $y$ and $z$ components of the particle canonical momentum $\mathbf{p}=m\mathbf{v}+q\mathbf{A}$
should be conserved, for each particle, for all time. This is a consequence
of the system Lagrangian $\mathcal{L}=m\mathbf{v}^{2}/2+q(\mathbf{v}\cdot\mathbf{A}-\phi)$
being independent of the $y$ and $z$ coordinates, as can be shown
from the Euler-Lagrange equations:
\[
\frac{d}{dt}\left(\frac{\partial\mathcal{L}}{\partial v_{q}}\right)=\frac{\partial\mathcal{L}}{\partial q}.
\]
The canonical momentum is defined as $\mathbf{p}=\frac{\partial\mathcal{L}}{\partial\mathbf{v}}$,
and hence is clear that for $q=y,z$:
\begin{equation}
\dot{p}_{y}=\dot{p}_{z}=0.\label{eq:cons-canonical-momenta}
\end{equation}

We seek to enforce this conservation property exactly. As we shall
see, this will constrain the form of the scattering of the magnetic
field to the particles in Eqs. \ref{eq:vx_p}-\ref{eq:vz_p}. Let's
focus on the conservation of $p_{y}$:
\begin{equation}
\dot{p}_{y}=m_{p}\dot{v}_{p,y}+q_{p}\dot{A}_{y,p}=0,\label{eq:py}
\end{equation}
where
\begin{equation}
A_{y,p}\equiv\sum_{i}A_{y,i}S_{l}(x_{p}-x_{i}).\label{eq:Ay-scatter-1}
\end{equation}
Equation \ref{eq:py} can be integrated over a substep $\nu$ to $\nu+1$,
to find (ignoring the subscript $y$): 
\begin{equation}
\left(m_{p}v_{p}+q_{p}A_{p}\right)^{\nu+1}-\left(m_{p}v_{p}+q_{p}A_{p}\right)^{\nu}=0,\label{eq:canonical_momenta_cons}
\end{equation}
which can be rearranged as :
\begin{equation}
\frac{v_{p}^{\nu+1}-v_{p}^{\nu}}{\Delta\tau^{\nu}}=-\frac{q_{p}}{m_{p}}\sum_{i}\frac{A_{i}^{\nu+1}S_{l}(x_{p}^{\nu+1}-x_{i})-A_{i}^{\nu}S_{l}(x_{p}^{\nu}-x_{i})}{\Delta\tau^{\nu}}\label{eq:discrete-py}
\end{equation}
Specializing this result for second-order splines ($l=2$), Taylor-expanding
the shape function, and casting Eq. \ref{eq:discrete-py} into the
form of Eq. \ref{eq:vy_p-1} gives (see App. \ref{app:B-field-scatter}),
\begin{equation}
B_{z,p}^{\nu+\nicefrac{1}{2}}=B_{0,z}+\sum_{i}\left[\frac{A_{y,i+1}^{\nu+\nicefrac{1}{2}}-A_{y,i}^{\nu+\nicefrac{1}{2}}}{\Delta x}S_{1}(x_{i+\nicefrac{1}{2}}-x_{p}^{\nu+\nicefrac{1}{2}})\right]+\frac{\Delta A_{y,i_{p}-1}^{\nu+\nicefrac{1}{2}}-2\Delta A_{y,i_{p}}^{\nu+\nicefrac{1}{2}}+\Delta A_{y,i_{p}+1}^{\nu+\nicefrac{1}{2}}}{8\Delta x^{2}}(x_{p}^{\nu+1}-x_{p}^{\nu}).\label{eq:Bzp}
\end{equation}
The first term on the right hand side is the central-difference approximation
of $B_{z}=\partial_{x}A_{y}$ at the particle location. In the second
term, $\Delta A_{y,i_{p}}^{\nu+\nicefrac{1}{2}}=A_{i_{p}}^{\nu+1}-A_{i_{p}}^{\nu}$.
The second term is an $O(\Delta\tau^{2})$ correction (commensurate
with the truncation error of the Crank-Nicolson scheme) evaluated
at the particle cell index $i_{p}$ that ensures exact conservation
of the particle canonical momentum. A similar procedure for the conservation
of $p_{z}$ yields: 
\begin{equation}
B_{y,p}^{\nu+\nicefrac{1}{2}}=B_{0,y}-\sum_{i}\left[\frac{A_{z,i+1}^{\nu+\nicefrac{1}{2}}-A_{z,i}^{\nu+\nicefrac{1}{2}}}{\Delta x}S_{1}(x_{i+\nicefrac{1}{2}}-x_{p}^{\nu+\nicefrac{1}{2}})\right]-\frac{\Delta A_{z,i_{p}-1}^{\nu+\nicefrac{1}{2}}-2\Delta A_{z,i_{p}}^{\nu+\nicefrac{1}{2}}+\Delta A_{z,i_{p}+1}^{\nu+\nicefrac{1}{2}}}{8\Delta x^{2}}(x_{p}^{\nu+1}-x_{p}^{\nu}).
\end{equation}
Note that, in 1D, $B_{x}$ must remain constant in space (because
$\nabla\cdot\mathbf{B}=\partial B_{x}/\partial x=0$) and time (because
$\partial_{t}B_{x}=\partial E_{z}(x)/\partial y-\partial E_{y}(x)/\partial z=0$).
The proposed scattering formula for the magnetic field components
along ignorable directions guarantees conservation of canonical momentum
for every particle sub-step. Conservation over the macro-step follows
straightforwardly by integration over all substeps.

\subsection{Binomial smoothing}

As in earlier studies \citep{birdsall-langdon,chen-jcp-11-ipic},
we apply binomial smoothing to reduce noise level of high $k$ modes
introduced by particle-grid interpolations \citep{birdsall-langdon}.
Smoothing preserves the conservation properties of the implicit Darwin
model when implemented appropriately. The governing Darwin-PIC equations
with binomial smoothing read:
\begin{eqnarray}
\epsilon_{0}\frac{E_{x,i+\nicefrac{1}{2}}^{n+1}-E_{x,i+\nicefrac{1}{2}}^{n}}{\Delta t}+SM(j_{x}^{n+\nicefrac{1}{2}})_{i+\nicefrac{1}{2}}-\left\langle j_{x}\right\rangle  & = & 0,\label{eq:ampere-final}\\
\frac{1}{\mu_{0}}\left.\partial_{x}^{2}\frac{A_{y}^{n+1}+A_{y}^{n}}{2}\right|_{i}+SM(j_{y}^{n+\nicefrac{1}{2}})_{i}-\left\langle j_{y}\right\rangle  & = & 0,\label{eq:vecpot_y-final}\\
\frac{1}{\mu_{0}}\left.\partial_{x}^{2}\frac{A_{z}^{n+1}+A_{z}^{n}}{2}\right|_{i}+SM(j_{z}^{n+\nicefrac{1}{2}})_{i}-\left\langle j_{z}\right\rangle  & = & 0.\label{eq:vecpot_z-final}\\
\frac{x_{p}^{\nu+1}-x_{p}^{\nu}}{\Delta t}-v_{x,p}^{\nu+\nicefrac{1}{2}} & = & 0,\label{eq:pcle-pos-final}\\
\frac{\mathbf{v}_{p}^{\nu+1}-\mathbf{v}_{p}^{\nu}}{\Delta t}-\frac{q_{p}}{m_{p}}\left(SM(\mathbf{E}^{n+\nicefrac{1}{2}})_{p}+\mathbf{v}_{p}^{\nu+\nicefrac{1}{2}}\times SM(\mathbf{B}^{\nu+\nicefrac{1}{2}})_{p}\right) & = & 0,\label{eq:pcle-v-final}
\end{eqnarray}
with the binomial operators $SM_{i}$ and $SM_{p}$ defined as: 
\begin{equation}
SM(Q)_{i}=\frac{Q_{i-1}+2Q_{i}+Q_{i+1}}{4},\label{eq:bi-nomial-smoothing}
\end{equation}
and 
\begin{equation}
SM(Q)_{p}=\sum_{i}SM(Q)_{i}S(x_{p}-x_{i}).
\end{equation}
Owing to the property in periodic domains that $\sum_{i}A_{i}SM(B)_{i}=\sum_{i}B_{i}SM(A)_{i}$,
it is straightfoward to show that energy and charge conservation theorems
remain valid \citep{chen-jcp-11-ipic}. Canonical momenta conservation
also survives when replacing $A$ by $SM(A)$ in the last section,
giving:
\begin{eqnarray*}
SM(B_{z}^{\nu+\nicefrac{1}{2}})_{p} & = & \sum_{i}\left[\frac{SM(A_{y}^{\nu+\nicefrac{1}{2}})_{i+1}-SM(A_{y}^{\nu+\nicefrac{1}{2}})_{i}}{\Delta x}S_{1}(x_{i+\nicefrac{1}{2}}-x_{p}^{\nu+\nicefrac{1}{2}})\right]+\\
 & + & \frac{SM(\Delta A_{y}^{\nu+\nicefrac{1}{2}})_{i_{p}-1}-SM(2\Delta A_{y}^{\nu+\nicefrac{1}{2}})_{i_{p}}+SM(\Delta A_{y}^{\nu+\nicefrac{1}{2}})_{i_{p}+1}}{8\Delta x^{2}}(x_{p}^{\nu+1}-x_{p}^{\nu}).
\end{eqnarray*}
i.e., $B_{z}$ must be scattered to particles from the binomially
smoothed $A_{y}$. A similar result is found for $SM(B_{y}^{\nu+1/2})_{p}$.

\section{Numerical tests\label{sec:numerical-tests}}

The set of field equations (\ref{eq:ampere-final}-\ref{eq:vecpot_z-final})
and particle equations (\ref{eq:pcle-pos-final}-\ref{eq:pcle-v-final})
are the ones solved in this study. For this, we employ a JFNK nonlinear
solver, implemented and configured as described in Ref. \citep{chen-jcp-11-ipic}.
As in the reference, the particle equations are enslaved to the field
equations (particle enslavement), which requires only a single copy
of the particle population. This results in minimal memory requirements
for the nonlinear solver, determined only by the storage required
by the field quantities.

In this section, we provide a sequence of numerical tests of increasing
complexity to provide verification against linear theory results (measured
as instability growth rates), and to demonstrate the favorable properties
of the approach. These tests are (from simplest to more complex):
an electron Weibel instability, an ion Weibel instability, and a kinetic
Alfv\'en wave problem. The first two tests are for non-magnetized
plasmas, and the last one is for a magnetized plasma. The last two
tests are stiff multiscale problems due to the ion-to-electron mass
disparity (all numerical tests employ a realistic mass ratio $m_{i}/m_{e}=1836$).

For these numerical tests, we normalize the Darwin PIC equations with
appropriate reference quantities:
\begin{equation}
\begin{array}{lll}
\hat{t}=t\omega_{0}, & \hat{x}=\frac{x}{x_{0}}, & \hat{v}=\frac{v}{v_{0}},\\
\hat{n}=\frac{n}{n_{0}}, & \hat{q}=\frac{q}{q_{0}}, & \hat{m}=\frac{m}{m_{0}},\\
\hat{E}=\frac{\varepsilon_{0}E}{q_{0}n_{0}x_{0}}, & \hat{A}=\frac{A}{\mu_{0}q_{0}n_{0}x_{0}^{3}}, & \hat{J}=\frac{J}{q_{0}n_{0}x_{0}\omega_{0}},
\end{array}
\end{equation}
to find:
\begin{eqnarray}
\frac{\partial\hat{E}_{x}}{\partial\hat{t}}+\hat{J}_{x} & = & \left\langle \hat{J}_{x}\right\rangle ,\\
\partial_{x}^{2}\hat{A}_{y,z}+\hat{j}_{y,z} & = & \left\langle \hat{j}_{y,z}\right\rangle ,\\
\hat{E}_{y,z} & = & -\frac{\partial\hat{A}_{y,z}}{\partial t},\\
\hat{B}_{y,z} & = & \left(\nabla\times\mathbf{\hat{A}}\right)_{y,z}\\
\frac{d\hat{x}_{p}}{d\hat{t}} & = & \hat{v}_{p},\\
\frac{d\hat{v}_{p}}{d\hat{t}} & = & \hat{a}_{p},
\end{eqnarray}
where $\hat{a}_{p}=\frac{\hat{q}_{p}}{\hat{m}}[\hat{E}_{p}+\hat{v}_{p}\times\hat{B}_{p}]$.
In a two-species system, we pick either electrons or ions as the reference
species, depending on the problem of interest. For electrons, the
associated reference constants are: 
\begin{eqnarray}
v_{0} & = & c\nonumber \\
x_{0} & = & d_{e}(=c/\omega_{pe}),\nonumber \\
\omega_{0} & = & \omega_{pe},\nonumber \\
n_{0} & = & \frac{W_{e}}{d_{e}},\\
q_{0} & = & \frac{n_{e}d_{e}e}{W_{e}},\nonumber \\
m_{0} & = & \frac{n_{e}d_{e}m_{e}}{W_{e}},\nonumber 
\end{eqnarray}
where $W_{e}=N_{e}/\hat{N}_{e}$ is the ratio of the number of real
particles and simulated particles. Similar reference values are found
for ions. Note that, in our units, the magnetic field reference value
is measured in units of $\hat{B}=\omega_{cs}/\omega_{ps}$, for $s=e$
or $i$.

\subsection{The electron Weibel instability}

The Weibel instability is an electromagnetic instability that can
appear in a unmagnetized plasma due to velocity-space anisotropy \citep{weibel1959spontaneously,fried1959mechanism}.
In a Cartesian coordinate system, a perturbation of the magnetic field
perpendicular to the wave vector $\boldsymbol{\mathbf{\mathit{k}}}$
(which is along the $x$ direction) can induce a plasma current that
increases the perturbation, provided that the plasma is hotter in
the perpendicular direction (i.e. $y$ and $z$). By making either
the electron or ion velocity distribution anisotropic, we can have
the electron or ion Weibel instability, respectively.

For the electron Weibel instability, we choose electrons as the reference
species. For the initialization of the particle distribution, we introduce
particles in pairs to obtain zero plasma current exactly for each
species. The two particles of each pair are set at the same location
with opposite velocities. The initial distribution function is 
\begin{equation}
f(x,v,t=0)=f_{M}(v)\left[1+a\cos\left(k_{x}x\right)\right]\label{eq:initf}
\end{equation}
where $f_{M}$ is the Maxwellian distribution, $a$ is the perturbation
level, $k_{x}$ is the perturbed wave number. The spatial distribution
is approximated by first putting ions randomly with a constant distribution,
e.g. $x_{0}\in[0,L]$. The electrons are distributed in pairs with
ions according to the Debye distribution \citep{williamson1971initial}.
Specifically, in each $e$-$i$ pair, the electron is situated away
from the ion by a small distance, $dx=\mathrm{ln}(R)$ where $R\in(0,1)$
is a uniform random number. The perturbation is done by shifting the
particle position by a small amount such that $x=x_{0}+a\cos\left(k_{x}x_{0}\right)$,
where $x_{0}$ is the initial particle position, $a=0.01$, and $k_{x}=\frac{2\pi}{L}$
with $L$ the domain size.

\begin{figure}
\begin{centering}
\includegraphics{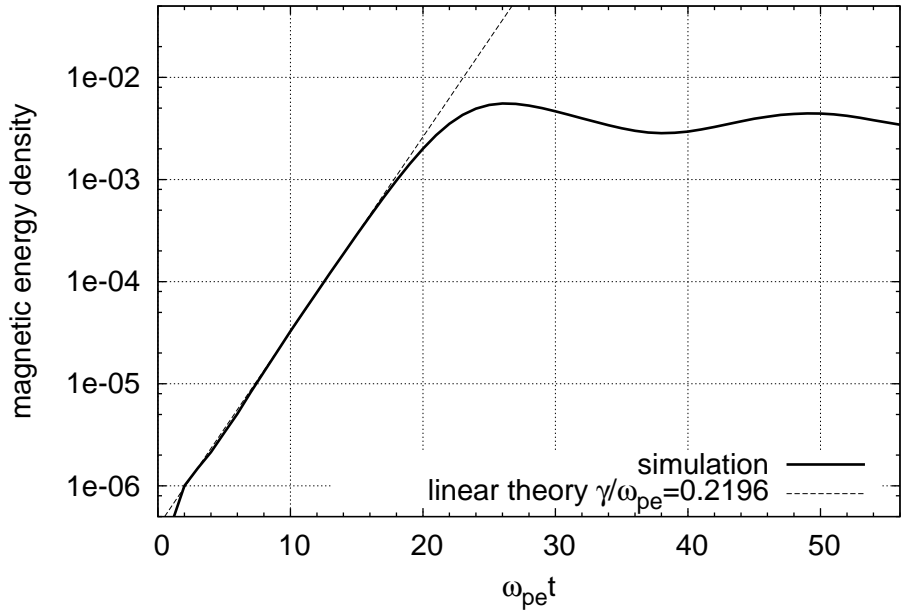}
\par\end{centering}

\caption{\label{fig:Weibel-instability}Time history of the magnetic field
energy $W_{A}=\sum_{i}(B_{y,i+\nicefrac{1}{2}}^{2}+B_{z,i+\nicefrac{1}{2}}^{2})/2$
evolving from an electron Weibel instability. Excellent agreement
with the theoretical linear growth rate is found.}
\end{figure}

\begin{figure}
\centering{}\includegraphics[scale=0.6]{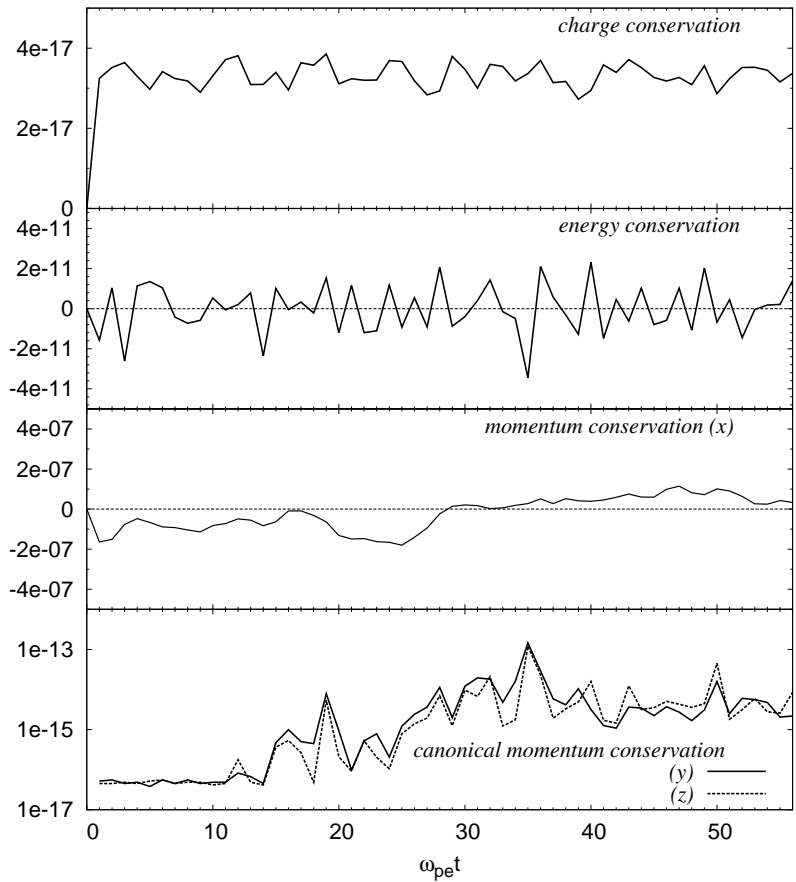}\caption{\label{fig:Weibel-conservations}Conserved quantities in the simulation
of the eletron Weibel instability. \textcolor{black}{Charge conservation
is measured as the (root-mean-square) rms of the continuity equation,
numerically evaluated at grid cells $\sqrt{\sum_{i}(\rho_{i}^{n+1}-\rho_{i}^{n}+\Delta t(\bar{j}_{i+\nicefrac{1}{2}}-\bar{j}_{i-\nicefrac{1}{2}})/\Delta x)^{2}/N_{g}}$
where $N_{g}$ is the number of grid-points. Energy conservation is
measured as the change in the total energy (c.f. Eq. \ref{eq:disc-totenergy})
between successive time steps. Momentum conservation in the $x$ direction
is measured as $\sum_{p}m_{p}v_{p,x}/\sum_{\alpha}mv_{th,x}$, where
$p$ and $\alpha$ indicate particle and species index respectively.
Finally, the maximum error in the conservation of canonical momenta
for all particles is measured as $\max_{p}\left(\mid m_{p}v_{p}^{n+1}+q_{p}A_{p}^{n+1}-m_{p}v_{p}^{n}-q_{p}A_{p}^{n}\mid\right)$
in the $y$ and $z$ directions, respectively.}}
\end{figure}

The plasma consists of electrons and singly charged ions, with a realistic
mass ratio $m_{i}/m_{e}=1836$. The simulated domain is of $\pi$
in length, with 64 uniform cells and periodic boundary conditions.
The average number of particles per cell of each species is 2000.
Electrons are initialized with an anisotropic Maxwell distribution
with $T_{ey,z}/T_{ex}=16$, and the thermal velocity parallel to the
wave vector is $v_{eTx}\equiv\sqrt{T_{ex}/m}=0.1$. Ions are initialized
with an isotropic Maxwell distribution with $v_{iTx}=0.1$. The timestep
is taken to be $\Delta t=1$. For comparison, the linear growth rate
($\gamma=0.22$) is found from the dispersion relation of electromagnetic
waves in a bi-Maxwellian plasma \citep{krall1973principles}:
\begin{equation}
1-\frac{k_{x}^{2}c^{2}}{\omega^{2}}-\sum_{\alpha}\frac{\omega_{p\alpha}^{2}}{\omega^{2}}\left(1+\frac{T_{\alpha y,z}}{2T_{\alpha x}}Z^{\prime}(\xi_{\alpha})\right)=0,
\end{equation}
where $\alpha=e,i$, $\xi_{\alpha}=\omega/k_{x}\sqrt{2T_{\alpha x}/m_{\alpha}}$,
and $Z^{\prime}(\xi)$ is the first derivative of plasma dispersion
function. The agreement between the simulation and theory is shown
in Fig. \ref{fig:Weibel-instability}. 
The time history of conserved quantities (e.g., charge, energy, momentum,
and canonical momenta) of the simulated system is depicted in Fig.
\ref{fig:Weibel-conservations}. We see that charge conservation is
at the computer round-off level. Energy conservation is determined
by the JFNK nonlinear tolerance level (a relative tolerance of $10^{-8}$
in used in this study), and the canonical momenta conservation is
determined by the Picard tolerance level for orbit integration (a
relative tolerance of $10^{-10}$ is used). As in earlier studies
\citep{chen-jcp-11-ipic}, the particle momentum in the $x$ direction
is not conserved exactly, but the error is relatively small. 

\subsection{The ion Weibel instability}

Next, we simulate the ion Weibel instability, which is more challenging
because electron dynamics makes the problem very stiff. We keep the
same mass ratio $m_{i}/m_{e}=1836$, but use ions as the reference
species for normalization. The simulated domain is of $\frac{2\pi}{3}\sqrt{m_{e}/m_{i}}$
in length, with 64 uniformly distributed cells (corresponding to a
cell width about 30 times lager than the Debye length), periodic boundary
conditions, and 2000 particles per cell of each species. The electron
species is initialized with an isotropic Maxwell distribution. We
consider two electron thermal velocities, $v_{eTx}=0.001$ and $v_{eTx}=0.025$.
The ion species is initialized with an anisotropic Maxwellian with
$T_{iy,z}/T_{ix}=40,000$ and $v_{iTx}=0.001$. The timestep is taken
to be $\Delta t=0.1\omega_{pi}^{-1}$, which is about a factor of
40 times larger than the Vlasov-Maxwell-PIC CFL. Relatively large
growth rates occur at large $k_{x}$ and large energy anisotropies,
consistent with those observed in Ref. \citep{chang1990electromagnetic}
(in which the anisotropy is introduced by a cross-field ion flow).
\begin{figure}
\begin{centering}
\includegraphics{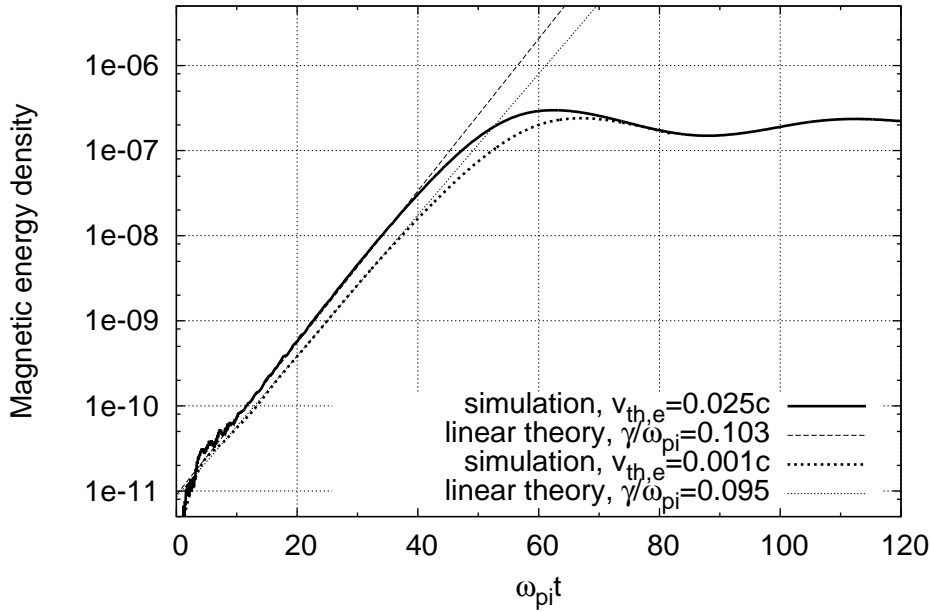}
\par\end{centering}

\caption{\label{fig:ion-Weibel-instability}Time history of the magnetic field
energy $W_{A}=\sum_{i}(B_{y,i+\nicefrac{1}{2}}^{2}+B_{z,i+\nicefrac{1}{2}}^{2})/2$
from the ion Weibel instability. Excellent agreement with the theoretical
growth rates for different electron thermal velocities is found.}
\end{figure}
Figure \ref{fig:ion-Weibel-instability} shows the time history of
the magnetic field energy density for the two electron thermal velocities.
Linear theory predicts growth rates of $9.5\times10^{-2}$ and $2.5\times10^{-2}$
for $v_{eTx}=0.001$ and $0.025$, respectively, which are in excellent
agreement with simulations.

\subsection{The kinetic Alfvén wave ion-ion streaming instability\label{par:The-kinetic-Alfven-case}}

\begin{figure}
\centering{}\includegraphics{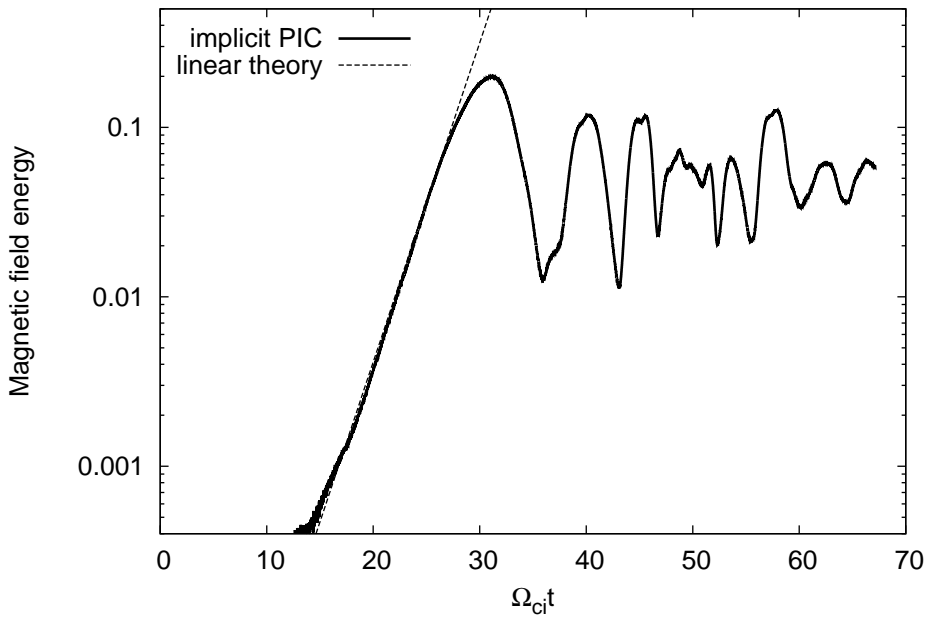}\caption{\label{fig:KAW}Time history of the magnetic field energy of the kinetic
Alfvén wave ion-ion streaming instability simulation, demonstrating
excellent agreement with linear theory. }
\end{figure}

Finally, we consider the excitation of kinetic Alfvén waves by ion-ion
streaming \citep{yin2007kinetic}. The instability is caused by interactions
between the wave and the streaming ions. The simulation parameters
are chosen to be similar to those presented in Ref. \citep{yin2007kinetic}.
The mass ratio is $m_{i}/m_{e}=1836$. We use ions as the reference
species. The simulated domain is $\frac{4\pi}{3}$ in length, with
64 uniformly distributed cells (with each cell about 40 times larger
than the Debye length) and periodic boundary conditions, and the average
number of particles per cell of one species is 2000. The external
magnetic field $B_{0}=0.00778$ is set to be at a large angle $\theta=70{}^{\circ}$
with respect to the propagation direction $(x)$ of the wave. The
plasma consists of Maxwellian electrons with $v_{eT}=0.0745$ ($\beta_{e}=0.1)$,
and two singly charged ion components, i.e., an ambient ion component
$a$ and an ion beam component $b$, with number densities $n_{a}=0.6n_{e}$
and $n_{b}=0.4n_{e}$ (where $n_{e}$ is the electron density). The
two ion components have $v_{aT}=0.0192$ and $v_{bT}=0.0745$, and
a relative streaming speed with respect to each other of $v_{d}=2.5v_{A}$,
with $v_{A}=\sqrt{m_{e}/m_{i}}/3$ the Alfvén speed along the external
magnetic field direction. The timestep is again set to $\Delta t=0.1\omega_{pi}^{-1}$
(about 20 times larger than the explicit CFL). Figure \ref{fig:KAW}
shows the simulation result of the magnetic energy density, which
is again in excellent agreement with linear theory (the growth rate
for this configuration is reported in Ref. \citep{yin2007kinetic}
to be $\gamma=0.218\omega_{pi}/\omega_{ci}$). 

\section{Discussion and conclusions\label{sec:conclusions}}

This study introduces a fully implicit Darwin-PIC algorithm that employs
a time-space-centered finite difference scheme for the coupled Darwin
field and particle equations. The non-radiative limit of Maxwell's
equations is of interest in non-relativistic regimes to avoid radiative
aliasing noise and/or instabilities \citep{langdon1972some,godfrey1974numerical},
particularly in the context of exactly energy conserving schemes \citep{markidis2011energy}.
We have used a potential formulation of the Darwin field equations,
in terms of vector potential $\mathbf{A}$ and electrostatic potential
$\phi$ (or equivalently $E_{x}$ in the 1D case), and standard Lagrangian
particle equations of motion (expressed in terms of position $\mathbf{x}$
and velocity $\mathbf{v}$). The stability of the algorithm is guaranteed
by the fully implicit nature of the scheme. In contrast to previous
Darwin-PIC algorithms \citep{birdsall-langdon,nielson-darwin-76},
the algorithm conserves global energy and local charge exactly in
the discrete. It also conserves particle canonical momenta in the
ignorable directions exactly, by carefully prescribing the magnetic
field scattering formula. A necessary condition for the energy conservation
is the exact reversibility of the time difference scheme, which is
guaranteed by our time-centered implicit discretization. Just as in
the electrostatic case \citep{chen-jcp-11-ipic}, charge conservation
is achieved by forcing particles to stop at cell boundaries as they
traverse their orbits, and by using first-order splines to gather
the current density. Orbit-averaging and binomial smoothing are introduced
without breaking the conservation properties of the scheme. Challenging,
stiff multiscale numerical tests have demonstrated the advertised
properties of the scheme, and its ability to employ large time steps
and cell sizes stably.

As in the electrostatic case \citep{chen2013fluid}, the ability of
the fully implicit Darwin-PIC approach to use large time steps and
cell sizes indicates much potential for algorithmic acceleration vs.
explicit Maxwell-PIC schemes (explicit Darwin-PIC implementations
are not available for such a comparison). Since the CFL condition
of explicit EM-PIC schemes (determined by the light speed) is more
stringent than that of explicit ES-PIC (determined by the fastest
thermal speed), we expect larger CPU speedups for implicit Darwin-PIC
than ES-PIC for comparable simulation parameters. We also expect the
convergence properties of the nonlinear solver to play a critical
role in the overall efficiency of the implicit Darwin-PIC algorithm.
Both of these are confirmed by the following back-of-the-envelope
analysis, which closely follows that in Ref. \citep{chen-jcp-11-ipic}
for ES-PIC (recently confirmed numerically in Ref. \citep{chen2013fluid}).

We begin by estimating the CPU cost for a given PIC solver to advance
the solution for a given time span $\Delta T$ as \citep{chen-jcp-11-ipic}:
\begin{equation}
CPU=\frac{\Delta T}{\Delta t}N_{pc}\left(\frac{L}{\Delta x}\right)^{d}C,\label{eq:CPU-estimation-1}
\end{equation}
where $N_{pc}$ is the number of particles per cell, ($L/\Delta x$)
is the number of cells per dimension, $d$ is the number of physical
dimensions, and $C$ is the computational complexity of the solver
employed, measured in units of a standard explicit PIC Vlasov-Maxwell
leap-frog timestep. Accordingly, the implicit-to-explicit speedup
is given by:
\[
\frac{CPU_{ex}}{CPU_{im}}\sim\left(\frac{\Delta x_{im}}{\Delta x_{ex}}\right)^{d}\left(\frac{\Delta t}{\Delta t_{ex}}\right)\frac{C_{ex}}{C_{im}},
\]
where we have assumed the same $N_{pc}$ for both explicit and implicit
schemes, and we denote $\Delta t$ to be the implicit timestep. For
simplicity, we assume that all particles take a fixed sub-timestep
$\Delta\tau$ in the implicit scheme, and that the cost of one time
step with the explicit PIC solver is comparable to that of a single
implicit sub-step. It follows that $C_{im}/C_{ex}\sim N_{FE}\left(\Delta t/\Delta\tau_{im}\right)$,
i.e., the cost of the implicit solver exceeds that of the explicit
solver by the number of function evaluations ($N_{FE}$, which is
a measure of the number of orbit evaluations) per $\Delta t$ multiplied
by the number of particle sub-steps ($\Delta t/\Delta\tau_{im}$,
a measure of the cost per orbit). As in earlier studies \citep{chen-jcp-11-ipic,chen2013fluid},
we consider an implicit time step comparable to ion time scales, i.e.
$\Delta t\sim\omega_{pi}^{-1}$. Assuming typical values for $\Delta\tau_{im}\sim\min[0.1\Delta x_{im}/v_{th},\omega_{ce}^{-1},\omega_{pi}^{-1}]$,
$\Delta t_{ex}\sim\Delta x_{ex}/c$, $\Delta x_{ex}\sim\lambda_{D}$,
$\Delta t\sim\omega_{pi}^{-1}$ and $\Delta x_{im}\sim0.2/k$, we
find that the CPU speedup scales as:
\begin{equation}
\frac{CPU_{ex}}{CPU_{imp}}\sim\frac{0.2}{(5k\lambda_{D})^{d}}\frac{c}{v_{th,e}}\min\left[\frac{1}{k\lambda_{D}},\frac{c}{v_{A}}\sqrt{\frac{m_{i}}{m_{e}}},\sqrt{\frac{m_{i}}{m_{e}}}\right]\frac{1}{N_{FE}},\label{eq:CPU-ex-im-1}
\end{equation}
where $v_{A}=B_{0}/\sqrt{n_{e}m_{i}\mu_{0}}$ is the Alfvén speed.
Compared with the ES case \citep{chen2013fluid}, the EM CPU speedup
is larger by a factor of $c/v_{th,e}$, as expected. As in the ES
case, Eq. \ref{eq:CPU-ex-im-1} confirms that the CPU speedup is inversely
proportional to $N_{FE}$. This motivates future work towards the
development of suitable fluid preconditioning strategies, as was done
in Ref. \citep{chen2013fluid} for the electrostatic case.

Finally, we acknowledge that the extension of the 1D-3V implicit Darwin-PIC
formulation to multiple dimensions is not straightforward, particularly
given the challenges documented in the literature \citep{nielson-darwin-76,hewett1994low}.
A main roadblock described in these studies is related to the enforcement
of the solenoidal constraint of the vector potential (or rather, the
complications stemming from its avoidance, particularly in regards
to boundary condition specification for the transverse component of
the electric field). In this regard, the Vlasov-Darwin formulation
considered in Sec. \ref{sec:Electromagnetic-Vlasov-Darwin} gives
us reason for optimism, since both the solenoidal constraint and Poisson's
equation are implicitly enforced in the continuum, and the transverse
component of the electric field can be readily found from the vector
potential. In the discrete, a necessary condition for the tractability
of this formulation is the ability to enforce exact local charge conservation
in multiple dimensions, which is within our reach \citep{chen-jcp-11-ipic,chen2013analytical}.
The implementation and demonstration of a multidimensional version
of our implicit Darwin-PIC algorithm will the subject of future work. 

\pagebreak{}

\appendix

\section{Time-preservation of the spatial average of a field satisfying Poisson's
equation in a 1D periodic domain}

\label{app:cons-spatial-average}

We demonstrate that a field $\xi(x,t)$ satisfying Poisson's equation,
\begin{equation}
\nabla^{2}\xi=S(x,t)\label{eq:Poisson}
\end{equation}
in a 1D periodic domain $[0,L]$ satisfies:
\begin{equation}
\partial_{t}\left\langle \xi\right\rangle =0,\label{eq:cons-average}
\end{equation}
with $\left\langle \cdots\right\rangle =\int_{0}^{L}dx[\cdots]$ the
spatial average. Note that $\left\langle S\right\rangle =0$ is a
solvability condition for the system \ref{eq:Poisson}, since $\left\langle \nabla^{2}\right\rangle =0$.

The formal proof begins by considering an augmented equation,
\begin{equation}
\partial_{t}\hat{\xi}=\frac{1}{\epsilon}\left[\nabla^{2}\hat{\xi}-S\right].\label{eq:augmented}
\end{equation}
The solution to this equation has the property that:
\begin{equation}
\lim_{\epsilon\rightarrow0}\hat{\xi}\rightarrow\xi,\label{eq:limit-property}
\end{equation}
i.e., $\xi$ is the quasi-static limit of $\hat{\xi}.$ Applying the
spatial average operator to Eq. \ref{eq:augmented}, and using the
solvability condition, we find:
\[
\partial_{t}\left\langle \hat{\xi}\right\rangle =0.
\]
Equation \ref{eq:cons-average} follows by taking the limit $\epsilon\rightarrow0$.

\section{Magnetic field scattering formulas for exact conservation of particle
canonical momenta\label{app:B-field-scatter}}

We begin with Eq. \ref{eq:discrete-py}, 
\begin{equation}
\frac{v_{p}^{\nu+1}-v_{p}^{\nu}}{\Delta\tau^{\nu}}=-\frac{q_{p}}{m_{p}}\sum_{i}\frac{A_{i}^{\nu+1}S_{l}(x_{p}^{\nu+1}-x_{i})-A_{i}^{\nu}S_{l}(x_{p}^{\nu}-x_{i})}{\Delta\tau^{\nu}}.\label{eq:discrete-py-1}
\end{equation}
We consider second-order splines ($l=2$). The analysis below can
be extended to higher-order splines, if needed, by keeping more terms
in the expansion. Taylor-expanding the $l=2$ shape functions at $x_{p,i}^{\nu+\nicefrac{1}{2}}\equiv(x_{p}^{\nu+\nicefrac{1}{2}}-x_{i})$,
we find:
\begin{eqnarray*}
S_{2}(x_{p}^{\nu+1}-x_{i}) & = & S_{2}(x_{p,i}^{\nu+\nicefrac{1}{2}})+(x_{p}^{\nu+1}-x_{p}^{\nu+\nicefrac{1}{2}})\left.\frac{\partial S_{2}}{\partial x_{p}}\right|_{x_{p,i}^{\nu+\nicefrac{1}{2}}}+\frac{(x_{p}^{\nu+1}-x_{p}^{\nu+\nicefrac{1}{2}})^{2}}{2}\left.\frac{\partial^{2}S_{2}}{\partial x_{p}^{2}}\right|_{x_{p,i}^{\nu+\nicefrac{1}{2}}},\\
S_{2}(x_{p}^{\nu}\:\:\:\:\:-x_{i}) & = & S_{2}(x_{p,i}^{\nu+\nicefrac{1}{2}})+(x_{p}^{\nu}\:\:\:\:\:-x_{p}^{\nu+\nicefrac{1}{2}})\left.\frac{\partial S_{2}}{\partial x_{p}}\right|_{x_{p,i}^{\nu+\nicefrac{1}{2}}}+\frac{(x_{p}^{\nu}\:\:\:\:\:-x_{p}^{\nu+\nicefrac{1}{2}})^{2}}{2}\left.\frac{\partial^{2}S_{2}}{\partial x_{p}^{2}}\right|_{x_{p,i}^{\nu+\nicefrac{1}{2}}}.
\end{eqnarray*}
No higher-order terms are present for $l=2$. Introducing these results
into Eq. \ref{eq:discrete-py-1}, we find:
\begin{eqnarray*}
\frac{v_{p}^{\nu+1}-v_{p}^{\nu}}{\Delta\tau} & = & \frac{q_{p}}{m_{p}}\sum_{i}\left[-\frac{A_{i}^{\nu+1}-A_{i}^{\nu}}{\Delta\tau}S_{2}(x_{p}^{\nu+\nicefrac{1}{2}}-x_{i})\right]\\
 & - & v_{x,p}^{\nu+\nicefrac{1}{2}}\sum_{i}\left[A_{i}^{\nu+\nicefrac{1}{2}}\left.\frac{\partial S_{2}}{\partial x_{p}}\right|_{x_{p,i}^{\nu+\nicefrac{1}{2}}}+\frac{A_{i}^{\nu+1}-A_{i}^{\nu}}{8}(x_{p}^{\nu+1}-x_{p}^{\nu})\left.\frac{\partial^{2}S_{2}}{\partial x_{p}^{2}}\right|_{x_{p,i}^{\nu+\nicefrac{1}{2}}}\right].
\end{eqnarray*}
Noting that, within a macro-step:
\[
-\frac{A_{i}^{\nu+1}-A_{i}^{\nu}}{\Delta\tau}=-\frac{A_{i}^{n+1}-A_{i}^{n}}{\Delta t}=E_{i},
\]
and comparing the velocity update above with Eq. \ref{eq:vy_p} (discretized
at $\nu+\nicefrac{1}{2}$), the definition of the magnetic field at
the particle position follows as: 
\begin{equation}
B_{z,p}^{\nu+\nicefrac{1}{2}}\equiv\sum_{i}\left[A_{y,i}^{\nu+\nicefrac{1}{2}}\left.\frac{\partial S_{2}}{\partial x_{p}}\right|_{x_{p,i}^{\nu+\nicefrac{1}{2}}}+\frac{A_{y,i}^{\nu+1}-A_{y,i}^{\nu}}{8}(x_{p}^{\nu+1}-x_{p}^{\nu})\left.\frac{\partial^{2}S_{2}}{\partial x_{p}^{2}}\right|_{x_{p,i}^{\nu+\nicefrac{1}{2}}}\right].\label{eq:bp-def}
\end{equation}
Here \citep{birdsall-langdon}:
\begin{eqnarray*}
\left.\frac{\partial S_{2}}{\partial x_{p}}\right|_{x_{p,i}^{\nu+\nicefrac{1}{2}}}=-\left.\frac{\partial S_{2}}{\partial x}\right|_{x_{p,i}^{\nu+\nicefrac{1}{2}}} & = & -\frac{S_{1}(x_{i+\nicefrac{1}{2}}-x_{p}^{\nu+\nicefrac{1}{2}})-S_{1}(x_{i-\nicefrac{1}{2}}-x_{p}^{\nu+\nicefrac{1}{2}})}{\Delta x},\\
\left.\frac{\partial^{2}S_{2}}{\partial x_{p}^{2}}\right|_{x_{p,i}^{\nu+\nicefrac{1}{2}}}=\left.\frac{\partial S_{2}^{2}}{\partial x^{2}}\right|_{x_{p,i}^{\nu+\nicefrac{1}{2}}} & = & \left\{ \begin{array}{cc}
1, & i=i_{p}-1\\
-2, & i=i_{p}\;\;\;\;\;\;\\
1, & i=i_{p}+1\\
0, & else
\end{array}\right.,
\end{eqnarray*}
with $i_{p}$ indicating the cell location of particle $p$. With
periodic boundary conditions, the first term on the right hand side
can be written as: 
\[
\sum_{i}A_{y,i}^{\nu+\nicefrac{1}{2}}\left.\frac{\partial S_{2}}{\partial x_{p}}\right|_{x_{p,i}^{\nu+\nicefrac{1}{2}}}=\sum_{i}\left[\frac{A_{y,i+1}^{\nu+\nicefrac{1}{2}}-A_{y,i}^{\nu+\nicefrac{1}{2}}}{\Delta x}S_{1}(x_{i+\nicefrac{1}{2}}-x_{p}^{\nu+\nicefrac{1}{2}})\right],
\]
which corresponds to the standard scattering formula for the magnetic
field at the particle position from a vector potential. The second
term on the right-hand-side of Eq. \ref{eq:bp-def} can be written
as: 
\[
\sum_{i}\frac{A_{y,i}^{\nu+1}-A_{y,i}^{\nu}}{8}(x_{p}^{\nu+1}-x_{p}^{\nu})\left.\frac{\partial^{2}S_{2}}{\partial x_{p}^{2}}\right|_{x_{p,i}^{\nu+\nicefrac{1}{2}}}=\frac{\Delta A_{y,i_{p}-1}^{\nu+\nicefrac{1}{2}}-2\Delta A_{y,i_{p}}^{\nu+\nicefrac{1}{2}}+\Delta A_{y,i_{p}+1}^{\nu+\nicefrac{1}{2}}}{8\Delta x^{2}}(x_{p}^{\nu+1}-x_{p}^{\nu}),
\]
where $\Delta A^{\nu+\nicefrac{1}{2}}=A^{\nu+1}-A^{\nu}$. This additional
term is a truncation error correction of $\mathcal{O}\left[(\Delta\tau^{\nu})^{2}\right]$,
which ensures exact canonical conservation for second-order shape
functions ($l=2$). A similar prescription can be found for $B_{y,p}$
from the conservation of $p_{z}$:
\[
B_{y,p}^{\nu+\nicefrac{1}{2}}=-\sum_{i}\left[A_{z,i}^{\nu+\nicefrac{1}{2}}\left.\frac{\partial S_{2}}{\partial x_{p}}\right|_{x_{p,i}^{\nu+\nicefrac{1}{2}}}\right]-\frac{\Delta A_{z,i_{p}-1}^{\nu+\nicefrac{1}{2}}-2\Delta A_{z,i_{p}}^{\nu+\nicefrac{1}{2}}+\Delta A_{z,i_{p}+1}^{\nu+\nicefrac{1}{2}}}{8\Delta x^{2}}(x_{p}^{\nu+1}-x_{p}^{\nu}).
\]
In our context, since the vector potential is periodic, a constant
external magnetic field component (if it exists) cannot be captured,
and  needs to be added explicitly. This can be readily done by adding
the corresponding constant magnetic field components to the scattering
formulas above.

The contribution of the constant magnetic field to the canonical momenta
must also be explicitly taken into account when diagnosing their conservation
in a periodic domain. This can be done as follows. For a constant
magnetic field $\mathbf{B}_{0}=(B_{0x},B_{0y},B_{0z})$, the change
in $A_{y}$ and $A_{z}$ (recall $A_{x}$ must remain constant to
enforce $\nabla\cdot\mathbf{A}=0$) over a substep along a particle
orbit is given by: 
\[
A_{y,p}^{\nu+1}-A_{y,p}^{\nu}=B_{0z}\Delta x_{p}^{\nu}-B_{0x}\Delta z_{p}^{\nu}\,\,;\,\, A_{z,p}^{\nu+1}-A_{z,p}^{\nu}=B_{0x}\Delta y_{p}^{\nu}-B_{0y}\Delta x_{p}^{\nu},
\]
where $(\Delta x_{p}^{\nu},\Delta y_{p}^{\nu},\Delta z_{p}^{\nu})=(v_{x,p},v_{y,p},v_{z,p})^{\nu+1/2}\Delta\tau^{\nu}$.

\section*{Acknowledgments}

The authors would like to acknowledge useful conversations with D.
A. Knoll, W. Daughton, and the rest of CoCoMans team. This work was
sponsored by the Los Alamos National Laboratory (LANL) Directed Research
and Development Program. This work was performed under the auspices
of the National Nuclear Security Administration of the U.S. Department
of Energy at Los Alamos National Laboratory, managed by LANS, LLC
under contract DE-AC52-06NA25396.\pagebreak{}\bibliographystyle{ieeetr}
\bibliography{kinetic}

\end{document}